\documentclass[sigconf]{acmart}
\usepackage[utf8]{inputenc}
\usepackage[T1]{fontenc}
\usepackage[inline]{enumitem}
\usepackage{booktabs}
\usepackage{amsmath}
\usepackage{amsfonts}
\usepackage{amssymb}
\usepackage{enumitem}
\usepackage{mathbbol}
\usepackage{mathtools}
\usepackage{hyperref}
\usepackage{url}
\usepackage{nicefrac}
\usepackage{microtype}
\usepackage{natbib}
\usepackage{multirow}
\usepackage{graphicx}
\usepackage{subcaption}
\usepackage{etaremune}
\usepackage{xcolor}

\newcommand{\eg}{\emph{e.g.}}

\newcommand{\passthrough}[1]{}
\newcommand{\todo}[1]{\passthrough{\textcolor{ACMBlue}{\textbf{[TODO] #1}}}}
\newcommand{\bodo}[1]{{\passthrough{\color{blue}[BB thinks: {\emph{#1}}]}}}

\AtBeginDocument{%
  \providecommand\BibTeX{{%
    \normalfont B\kern-0.5em{\scshape i\kern-0.25em b}\kern-0.8em\TeX}}}

\setcopyright{acmcopyright}
\copyrightyear{2018}
\acmYear{2018}
\acmDOI{10.1145/1122445.1122456}
\acmConference[Woodstock '18]{Woodstock '18: ACM Symposium on Neural
  Gaze Detection}{June 03--05, 2018}{Woodstock, NY}
\acmBooktitle{Woodstock '18: ACM Symposium on Neural Gaze Detection,
  June 03--05, 2018, Woodstock, NY}
\acmPrice{15.00}
\acmISBN{978-1-4503-XXXX-X/18/06}

\begin{document}

\title{ORCAS: 18 Million Clicked Query-Document Pairs for~Analyzing~Search}

\author{Nick Craswell}
\email{nickcr@microsoft.com}
\affiliation{%
  \institution{Microsoft}
  \city{Bellevue}
  \country{USA}
}
\author{Daniel Campos}
\email{dacamp@microsoft.com}
\affiliation{%
 \institution{Microsoft, University of Washington}
  \city{Bellevue}
  \country{USA}
}
\author{Bhaskar Mitra}
\email{bmitra@microsoft.com}
\affiliation{%
  \institution{Microsoft, University College London}
  \city{Montr\'eal}
  \country{Canada}
}
\author{Emine Yilmaz}
\email{emine.yilmaz@ucl.ac.uk}
\affiliation{%
  \institution{University College London}
  \city{London}
  \country{UK}
}

\author{Bodo Billerbeck}
\email{bodob@microsoft.com}
\affiliation{%
  \institution{Microsoft}
  \city{Melbourne}
  \country{Australia}
}

\begin{abstract}

  Users of Web search engines reveal their information needs through queries and clicks, making click logs a useful asset for information retrieval. However, click logs have not been publicly released for academic use, because they can be too revealing of personally or commercially sensitive information. This paper describes a click data release related to the TREC Deep Learning Track document corpus. After aggregation and filtering, including a \emph{k}-anonymity requirement, we find 1.4 million of the TREC DL URLs have 18 million connections to 10 million distinct queries. Our dataset of these queries and connections to TREC documents is of similar size to proprietary datasets used in previous papers on query mining and ranking. We perform some preliminary experiments using the click data to augment the TREC DL training data, offering by comparison: 28x more queries, with 49x more connections to 4.4x more URLs in the corpus. We present a description of the dataset's generation process, characteristics, use in ranking and other potential uses.

\end{abstract}



\maketitle

\section{Introduction}
\label{sec:intro}

\todo{R1: position bias: Did the authors take that into consideration when they collected the data?}

\todo{R1: ORCAS doesn't contain impressions.}
\bodo{maybe we should explicitly mention that adding impression data would be too commercially valuable.}

\todo{R1: It can be the difficult in aligning TREC DL and ORCAS queries.}

\todo{R2: The authors do not clarify which dataset of MS MARCO was used. They say that the queries from the Deep Learning track are extracted from the question answering dataset, but they do not say anything about the documents.}

\todo{R2: The type of information that the dataset contains is not novel.}

\todo{R2: Although this resource release means a step that improves some of the problems of previous datasets, I am very skeptical about the impact this dataset may have on research results in the future.}


\todo{R2: Figure 3: The x-axis of the left image is messed up. I guess that it ranges from 0 to 20, but I am not sure.}

\todo{R2: Figure 3: I think that the range of the y-axis is not correct. If they are absolute values, the sum of each bar's value should sum up to the total number of queries (10.4M). It seems that they are percentages, but it is not explained anywhere.}

\todo{R3: The contribution goes some way towards addressing the lack of commercial-scale click logs available for Information Retrieval research purposes.  However understandably, there are for commercial reasons no user id’s, no session information, no document ranks, no negative examples, and no dwell time information – just clicks onto TREC docs.}

\todo{R3: One criticism I have is that I’m not sure that the experiments in Section 5 contribute much to the paper.  I was much more interested in reading about the analysis of the click-document graph and the re-evaluation of last years runs – and would have welcomed an expansion of this type of discussion vs. the presented retrieval runs.}

\todo{R3: The data is published for “non-commercial research purposes only”.  So, while IANAL, the dataset appears to be generally available for academic research purposes, but there seems to be a restriction on usage by industry practitioners.}

\todo{R3: Predicted Impact:  I think the immediate impact of this contribution will be contained within the TREC DL Track community.  Further, there may be impact from enabling research studies that pick up this data for query mining applications.}


When people search the Web, they reveal their information needs through their queries and clicks. Such click logs become an important asset. For example, given the ambiguous query \emph{panda}, click logs can tell us the relative popularity of different interpretations, such as animals, movies, songs, restaurants and technology products. We can ask: Which documents are clicked for the query \emph{panda}? Which topics are most popular in panda-related queries? What topics are most popular in panda-related documents? Popularity can also vary by context, for example Australia's PANDA Helpline is more commonly clicked by Australian users, but less so by US users.

Not many click datasets are publicly available, for two reasons. One is privacy. Users enter sensitive and private information into a search engine, so a query can reveal information that should not be shared. By analyzing a stream of queries from the same user, we may see them search for their own name, address, or other details. Using these, and other hints, we can discover a lot about a person \cite{barbaro2006face}. Serious problems are raised by the release of queries linked to user IDs, or even session IDs. Even without such links, it is dangerous to release a tail query that was only typed by one user.

The other barrier to sharing click data is the commercial value of the data. In a particular country or language, the search engine with the most search traffic has an advantage over smaller competitors, because it has more information about user needs. Sharing that data at scale would help competitors, including new entrants in the market, and potentially help search engine optimizers. It could also reveal information about the workings of the engine, for example if results lists were provided including rank positions of the clicked and unclicked results.

Given these barriers, one option is to provide anonymized click data, such as the Yandex search personalization challenge \cite{serdyukov2014log}. The 35 million search sessions in the dataset (a Kaggle competition with 194 teams) have anonymized user IDs, queries, query terms, URLs, URL domains and clicks. Providing these as numeric IDs rather than strings greatly reduces concerns about privacy and commercial value. It can't be used to build a competing search engine because it has URL IDs rather than URLs. It can't be used to identify a user who entered their name as a query, because we just have a query ID and some term IDs.

The main disadvantage of anonymization via IDs is that we can not add new relevance judgments, or apply deep models that bring to bear text embeddings. It is also not possible to for instance discuss the meaning of the term \emph{panda} because we do not know which termID maps to that term.

This paper describes a new dataset, the Open Resource for Click Analysis in Search (ORCAS). Rather than providing user-level or session-level information, which can be used for personal identification, we focus on query-document connections. We use queries that have been repeated across many users. ORCAS data has clicks for TREC documents only, and only for English-speaking users in the United States. Combined with the filtering and aggregation, the dataset is intended to provide a useful resource for ranking TREC documents and also Web search mining more generally, without releasing personally identifying or commercially valuable information.

\section{Related work}
\label{sec:related}

\begin{table*}[]
\centering
\caption{Size comparison of query-URL pair datasets used in Web mining and ranking studies. We leave a blank if a size was not provided, and use the filtered size in cases where filtering was applied before use. Our new dataset is larger than several of the previous datasets, suggesting those past results could be replicated or extended using ORCAS.}
\begin{tabular}{llllll}
\toprule
Paper    & Queries & URLs & Q-U pairs & Availability & Primary focus of paper \\
\midrule
\citet{Beeferman:etal:KDD2000} & 244K & 362K & 1.9M & proprietary & Related Q \\
\citet{Wen01} & 1M/week & Encarta & 500K/week & proprietary & Related Q \\
\citet{Xue04} & 862K & 507K & & proprietary & Ranking \\
\citet{Craswell07} & 202K & 505K & 1.1M & proprietary & Ranking \\
\citet{Baeza-Yates07} & 7.5M & 973K & & proprietary & Related Q \\
\citet{Mei08} & 637M & 585M & & proprietary & Related Q \\
\citet{huang2013learning} &  &  & 100M & proprietary & Ranking \\
\midrule
ORCAS & 10.4M & 1.4M & 18.8M & open & Ranking \\ 
TREC DL \cite{craswell2020overview}& 367K & 320K & 384K & open & Ranking \\
\bottomrule
\end{tabular}
\label{tab:click_graph_sizes}
\end{table*}

There have been several previous attempts at releasing click datasets. The AOL data~\cite{AOL06} and MSN data~\cite{zhang2006some} were two of the initial attempts towards releasing click data, which came with question marks over their use. 

The AOL dataset contained $\sim 20$ million queries together with the domains of associated clicked URLs from $\sim 659,000$ users, collected over a 3-month period from March to May 2006. It was found to allow personal identification of some individual people \cite{barbaro2006face}, which was a major setback for the release of any future similar datasets. As a result of this, the dataset was retracted and is no longer available.  

The MSN data~\cite{zhang2006some} contained $\sim 15$ million queries that were collected from the users of the MSN search engine from the United States. The dataset was released in 2006 and again in 2009, but the agreement required a limited-time use, usage has now timed out, and it has not been released again since 2009. 

Some years later, as part of the Yandex search personalization challenge, Yandex released a dataset with $\sim 35$ million search sessions~\cite{serdyukov2014log}. The dataset was created using search activity of users over 27 days as the training data, with a separate test data containing user activity over 3 days. As mentioned in the previous section, this dataset contains URL IDs as opposed to the actual URL itself, which limits the type of research that can be conducted with the dataset.

The Sogou dataset~\footnote{\url{http://www.sogou.com/labs/}}, released by the major Chinese search engine company Sogou, contains $\sim 25.1$ million queries and $\sim 43.5$ million user interactions in the form of submitted queries and clicked Web page search results.
Almost all queries in this dataset contain solely characters from the simplified Chinese alphabet – which has around 3,500 commonly used characters~\cite{Whiting14}.

Since the aforementioned datasets have problems related to availability and use cases, a significant amount of research has been conducted on proprietary datasets instead. Proprietary datasets containing queries with associated clicked URLs have widely been used for ranking purposes ~\cite{Craswell07, Xue04, huang2013learning}, as well as for identifying related queries~\cite{Mei08, Baeza-Yates07, Wen01, Beeferman00, mitraexploring}. However, such proprietary datasets are not made available to the wider research community, which has been a major limitation for research.

The TREC Deep Learning (DL) Track~\cite{craswell2020overview} attempted to address the need for large amounts of training data by releasing large scale training datasets that are based on human relevance assessments, derived from MS MARCO~\cite{bajaj2016ms}. The track focuses on the document retrieval and passage retrieval tasks. In the document retrieval task,
367,013
queries together with associated relevance labels were made available, corresponding to a corpus of $3.2$ million documents. The training datasets released as part of the track are sparse, with no negative labels and often only one positive label per query, analogous to some real-world training data such as click logs. 

While the Deep Learning Track is a step forward towards making a large scale dataset that can be used for training information retrieval systems publicly available, the datasets released by the track are still based on human relevance assessments as opposed to real usage. The ORCAS dataset released as part of this paper can be seen as complementary to the data released as part of the Deep Learning Track since it is based on click logs.

Table~\ref{tab:click_graph_sizes} shows a comparison of different proprietary datasets, datasets released as part of the TREC DL Track, and the ORCAS dataset
in terms of size (number of queries, URLs and query-URL pairs) and availability of the datasets, as well as the application areas for which the datasets have primarily been created for.  

\section{Dataset collection}
\label{sec:data}

\begin{table*}[]
\centering
\caption{Summary of ORCAS data. Each record in the main file (\texttt{orcas.tsv}) indicates a click between a query (Q) and a URL (U), also listing a query ID (QID) and the corresponding TREC document ID (DID). For use in ranker training, we also provide files in TREC format for qrels, queries and runs. The run file is the top-100 using Indri query likelihood, for use as negative samples during training.}
\begin{tabular}{lrrl}
\toprule
\toprule
Filename & Size & Records & Data in each record \\
\midrule
\texttt{orcas.tsv} & 1.76GB & 18.8M & \texttt{QID Q DID U}\\
\texttt{orcas-doctrain-qrels.tsv} & 410MB & 18.8M & \texttt{QID DID} \\
\texttt{orcas-doctrain-queries.tsv} & 322MB & 10.4M & \texttt{QID Q}\\ 
\texttt{orcas-doctrain-top100} & 52.8GB & 983M & \texttt{QID DID score} \\
\bottomrule
\end{tabular}
\label{tab:data_size}
\end{table*}

\begin{figure}
    \centering
    \includegraphics[width=0.7\linewidth]{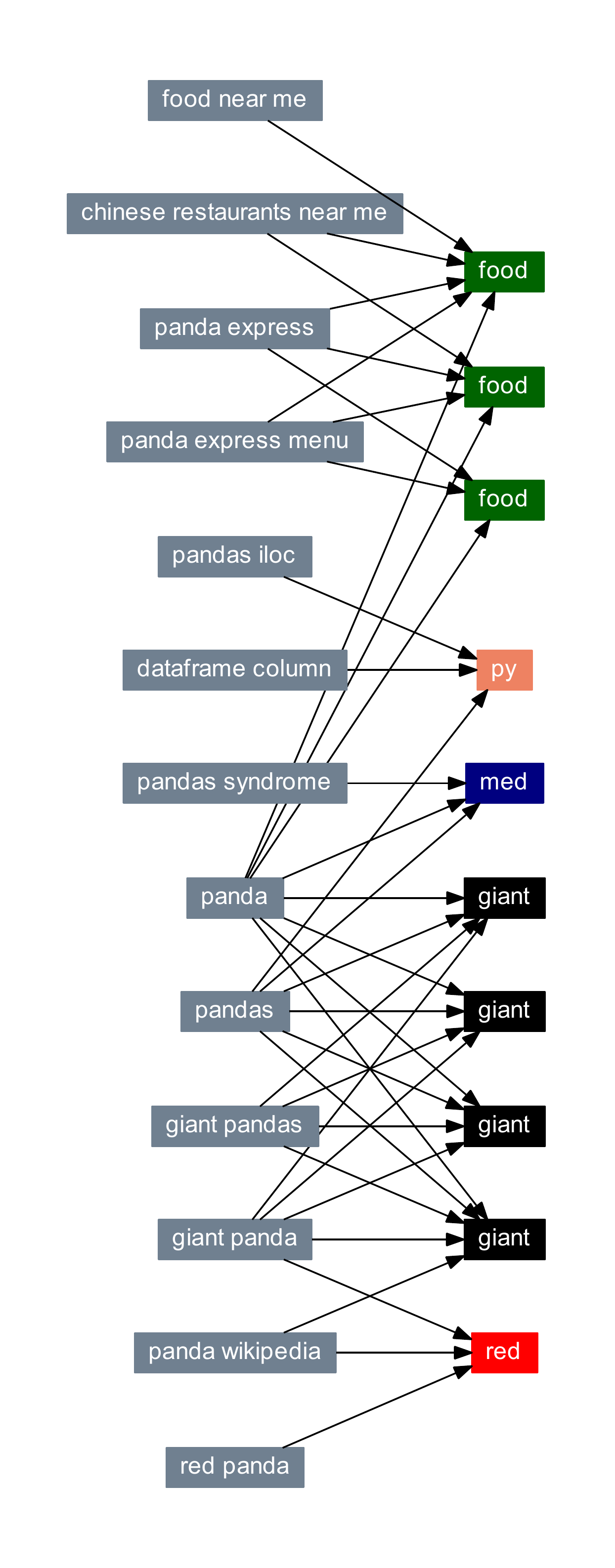}
    \caption{Example click data relating to the string `panda'. Query nodes are on the left and URL nodes are on the right. A query-URL connection means an edge exists in ORCAS data. The topics of the URLs are: \emph{food} Panda Express, \emph{py} Pandas Python library, \emph{med} Pandas syndrome, \emph{giant} Giant pandas, and \emph{red} Red pandas}
    \label{fig:clickgraph}
\end{figure}

The ORCAS dataset was created through log mining, aggregation and filtering.

Search engine logs can include information about the user's query, the ranking of URLs, the URLs that were clicked, as well as post-click behavior such as dwell time \cite{hofmann2016online}. We focus on click events, since in aggregate these are a good positive signal for relevance. Before aggregation we apply some minimal filtering, to eliminate clicks that had very negative post-click signals, such as a short dwell, since this can indicate that the user was disappointed with the result. 

We aggregate click events at the query-URL level, without considering context such as other URLs or their rank order. 
For the purpose of collating this dataset, we filter for users that are located in the United States and speak English. But we do not collect any other data about the users such as their search session or any other short term or long term personalizing information \cite{Bennett:2012}.
Such things could potentially be studied using the anonymized Yandex personalization challenge data \cite{serdyukov2014log}, but not using our aggregation to the query-URL level. Furthermore, rather than counting how popular query-URL pairs are during aggregation, we simply note which query-URL pairs are present, to avoid revealing too much information about event popularity in our logs.
We will describe in Section~\ref{sec:analysis} how it is possible to recover some information about popularity, even though we have removed the per-pair popularity information.

The full set of query-URL data, aggregated based on a subsample of Bing's 26-month logs to January 2020, would still be too commercially valuable, potentially covering billions of queries. It could also potentially reveal information known only to one person. We apply several strict filters.
First, keep only query-URL pairs where the URL is present in the 3.2 million document TREC DL corpus. We needed to choose some small subset of all clicked documents, so aligning with a TREC seems like a suitable choice. Second, we apply a \emph{k}-anonymity filter, keeping only queries that were typed by \emph{k} different users, for a high value of k. This makes it impossible for our dataset to contain a query with information that is only known to fewer than \emph{k} users. Finally, we applied filters to remove potentially offensive queries, for example queries related to pornography or hate speech. 

Two of our design decisions relating to ranking and position bias can now be described in more detail in relation to our \emph{k}-anonymity requirement.

Firstly, although ORCAS includes clicked URLs, it does not provide the rankings of clicked and unclicked URLs that users saw. If released without an anonymity requirement, such ranking information could be too commercially valuable, revealing information about query popularity, ranking variation and user preferences. With a high enough \emph{k} threshold, as used here, the data size would be reduced by orders of magnitude. The requirement is that the same ranking was seen by \emph{k} users who all selected the same URL. Because rankings vary from user to user and also vary over time, very few cases reach the \emph{k} threshold.

Secondly, although it would be possible to correct for position bias before aggregation, ORCAS does not do this. Correcting for position bias would not remove any URLs from the dataset, because results with \emph{k} or more clicks tend to be relevant, independently of position. Instead, position bias correction could allow us to include lower-ranked documents in ORCAS, which had fewer clicks due to position but were equally relevant. However, for privacy reasons we can not include results clicked by fewer than \emph{k} users. Therefore we focus on clicks in aggregate, uncorrected by position bias. This approach, of using uncorrected clicks as a positive signal, has proven useful in a variety of studies such as those in Table~\ref{tab:click_graph_sizes}.

Our main dataset becomes 18.8 million records with four columns per record:
\begin{verbatim}
    QID: 10103699
    Q: why is the sky blue
    DID: D1968574
    U: http://www.sciencemadesimple.com/sky_blue.html
\end{verbatim}
The document ID (DID) is the same  as the one used in the TREC corpus. To avoid revealing the unreleased query IDs in our held out test set, we assigned a disjoint set of query IDs (QIDs) to the ORCAS queries. This means the same query string can occur in TREC DL and ORCAS data, with different QIDs.

We also provide the same data in TREC format, as qrels, queries and Indri rankings for use in negative sampling (see Section~\ref{sec:experiment}). These comprise the full ORCAS data release\footnote{\url{https://microsoft.github.io/TREC-2020-Deep-Learning/ORCAS}}, as described in Table~\ref{tab:data_size}.

\subsection*{Dataset examples}

Figure~\ref{fig:clickgraph} shows a sample of the ORCAS data, for some documents related to various meanings of the term \emph{panda}. Some popular URLs are about Panda Express restaurants, the Python library Pandas, Pandas Syndrome, Giant Pandas or Red Pandas. Since URLs are too large to fit in the figure, we label each URL node with the type of panda it covers: food, py, med, giant and red. For example, one of the nodes labeled `giant' is the Wikipedia page \url{https://en.wikipedia.org/wiki/Giant_panda}. The queries were selected for being high-degree nodes, both in the global corpus and for this set of documents. A query such as \emph{pandas syndrome} is only connected to documents of one topic. Whereas a query like \emph{panda} is ambiguous, connected to documents of multiple topics. 

Absence of an edge can indicate that the document is less relevant, for example the Python Pandas node is not adjacent to \emph{panda}, because it is less likely that a user who typed that query wants Python Pandas. Edge absence can also indicate that the document or query is less popular, and also indicate specific patterns of retrieval in the underlying search engine. This may explain why the Red Panda URL is not connected to \emph{panda} or \emph{pandas}, despite being on-topic.

The figure covers several topics, but in the ORCAS dataset its possible to identify several more panda topics, such as Kung Fu Panda movies, the debut single of rapper Desiigner and Panda Antivirus. Outside the TREC documents and US click data used here there are even more meanings of panda, such as the Australian PANDA Helpline or the PandA gateway at Kyoto University. 

To illustrate the mining of related queries, in the following paragraph we give an example based on the seed query \emph{orcas} (QID=2126294).
Related queries can be found by taking a two-step walk. The first step reaches eight URLs. The second step reaches 198 queries. We can rank the queries according to the volume of paths, assuming the queries that are reachable by more distinct paths are more related to the seed query. The most similar queries (with 4--5 paths) are about whales: \emph{orca whale}, \emph{orca facts}, \emph{orca}, \emph{orca whales} and \emph{killer whale facts}. Somewhat less related (with 2 paths) are queries about Orcas Island: \emph{orcas islands}, \emph{orcas island washington}, \emph{orcas island wa} and \emph{orcas island}.



\section{Dataset analysis}
\label{sec:analysis}

\begin{figure}
    \centering
    \includegraphics[width=.48\linewidth]{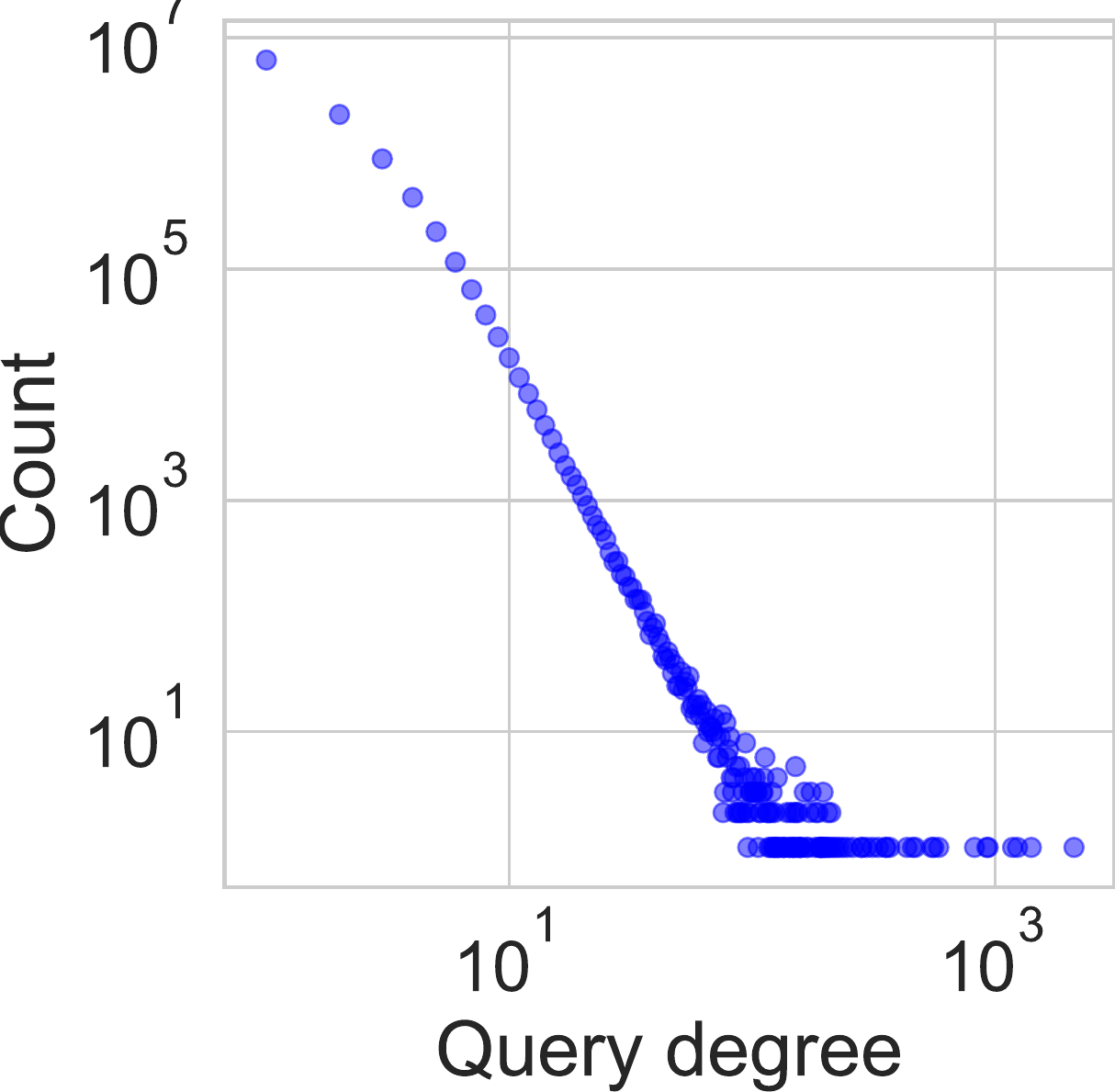}
    \includegraphics[width=.48\linewidth]{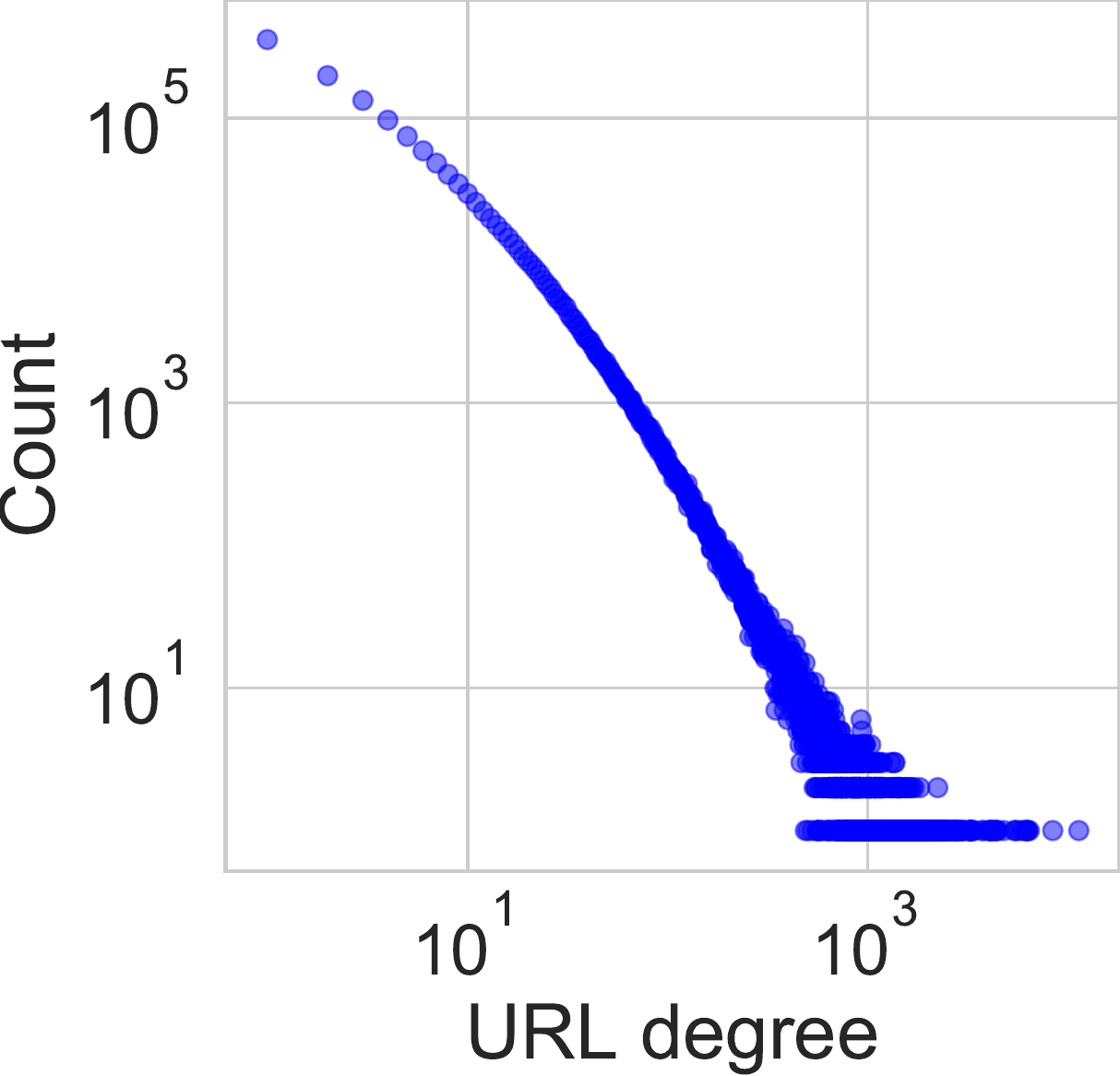}
    \caption{Degree distribution of query nodes and URL nodes in the bipartite query-URL graph.}
    \label{fig:degree_distribution}
\end{figure}

\begin{figure}
    \centering
    \includegraphics[width=.48\linewidth]{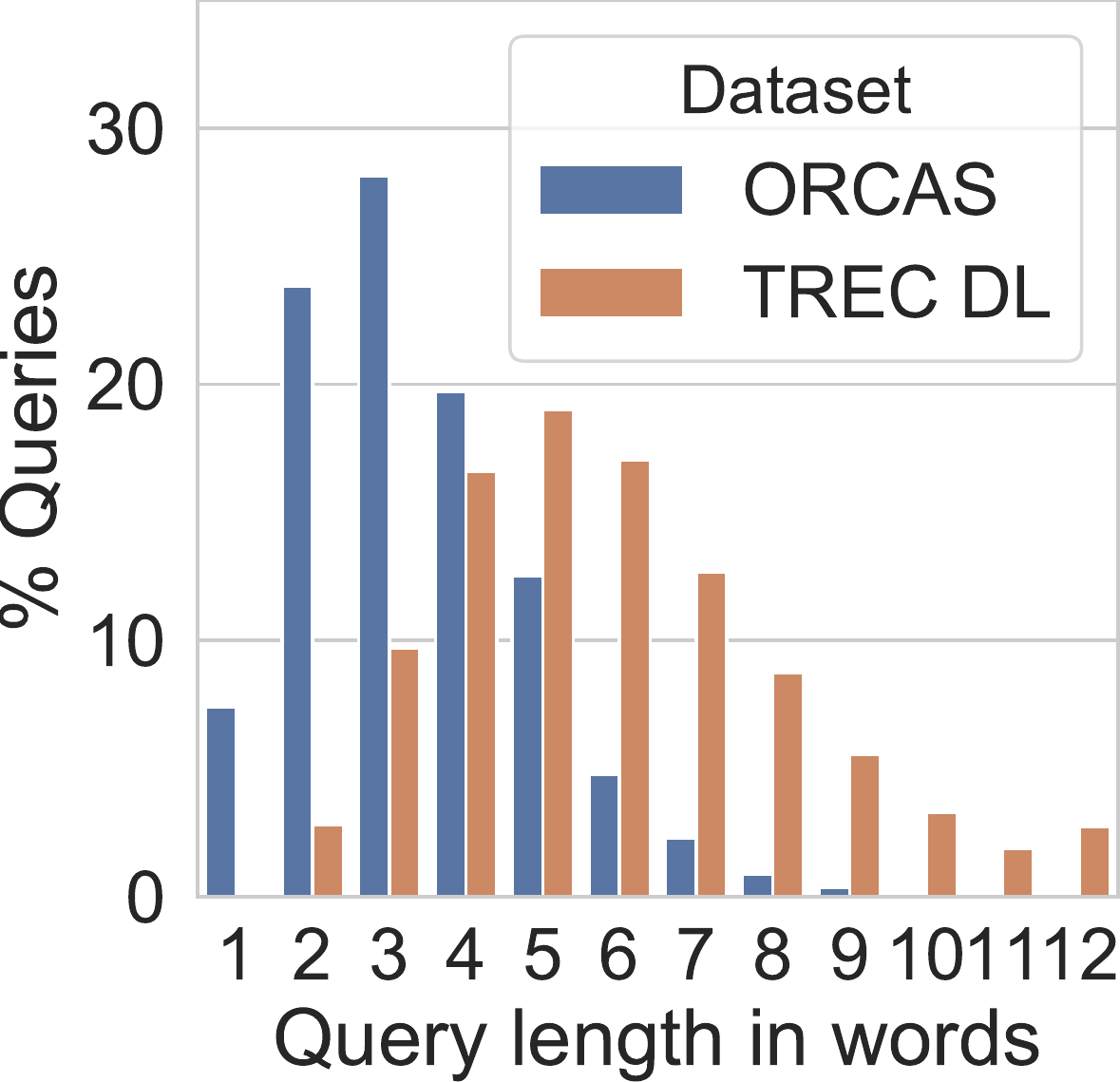}
    \includegraphics[width=.48\linewidth]{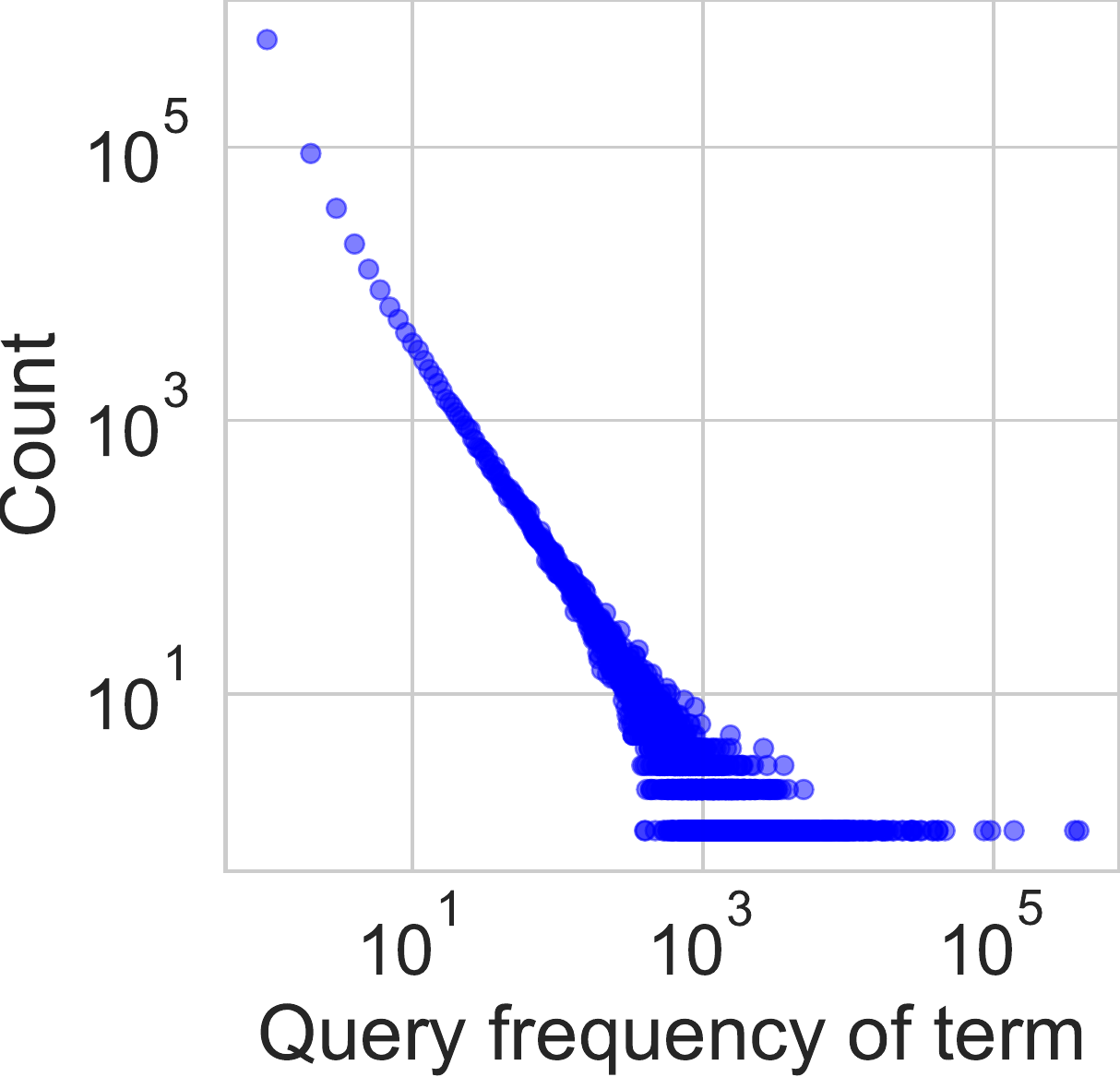}
    \caption{ORCAS queries are shorter than TREC DL queries. ORCAS queries contain a variety of query terms, from the most common term \emph{how} is in 387K queries, and 613K different terms that each occur in one query (for example \emph{egyptlive}).}
    \label{fig:orcas_terms}
\end{figure}

\begin{table}[]
\centering
\caption{The most common first word in TREC DL queries is \emph{what},  in 39.4\% of the queries (145K queries). The most common first word in ORCAS queries is \emph{how}, in 3.7\% of queries (387K queries). TREC DL queries were selected in a way that favored natural language questions, so have less variety of first words, and more focus on question words. ORCAS queries were selected based on the TREC DL documents, so still have question words, but also have words in the top-10 that are more rare in TREC DL such as \emph{www}, \emph{the}, \emph{best} and \emph{free}. }
\begin{tabular}{lrrrrrr}
\toprule
& \multicolumn{3}{c}{TREC DL} & \multicolumn{3}{c}{ORCAS} \\
\cmidrule(lr){2-4} \cmidrule(lr){5-7}
Word & Count & \% & [Rank] & Count & \% & [Rank]\\
\midrule
what & 144575 & 39.4\% & [1] & 362169 & 3.5\% & [2] \\
how & 50113 & 13.7\% & [2] & 386611 & 3.7\% & [1] \\
where & 16091 & 4.4\% & [3] & 46331 & 0.4\% & [6] \\
who & 12957 & 3.5\% & [4] & 38522 & 0.4\% & [9] \\
is & 10742 & 2.9\% & [5] & 41512 & 0.4\% & [8] \\
when & 10002 & 2.7\% & [6] & 37885 & 0.4\% & [10] \\
which & 6138 & 1.7\% & [7] & 7347 & 0.1\% & [89] \\
can & 5736 & 1.6\% & [8] & 20354 & 0.2\% & [21] \\
average & 5389 & 1.5\% & [9] & 9906 & 0.1\% & [52] \\
define & 4620 & 1.3\% & [10] & 27469 & 0.3\% & [14] \\
www & 20 & 0.0\% & [379] & 138528 & 1.3\% & [3] \\
the & 2289 & 0.6\% & [15] & 95685 & 0.9\% & [4] \\
best & 203 & 0.1\% & [58] & 86119 & 0.8\% & [5] \\
free & 16 & 0.0\% & [487] & 41514 & 0.4\% & [7] \\
\bottomrule
\end{tabular}
\label{tab:first word}
\end{table}

This section describes characteristics of the dataset and compares it to other datasets in the literature.


Query-URL datasets have been used for improving document ranking, finding related queries, and other log mining applications, as was indicated in Table~\ref{tab:click_graph_sizes}. For reasons of commercial sensitivity, ORCAS does not contain ``edge weights'', which could indicate the relative popularity of URLs for a query, or vice versa. However, it is still possible to recover some popularity information from ORCAS. Specifically, queries that are more popular tend to accumulate connections to a greater number of URLs, and URLs that are commonly clicked tend to accumulate connections to a greater number of queries. Popularity is indicated by higher degree in the bipartite query-URL graph. 

Figure~\ref{fig:degree_distribution} shows the degree distributions of nodes in the ORCAS bipartite graph. Skewed degree distributions, which are better viewed on a log-log scale such that a few nodes are very highly connected and many nodes have only one connection, are common for datasets of this sort. Very popular queries
(such as~\emph{weather})
have hundreds of URLs, while very popular URLs
(such as \url{www.outlook.com})
have thousands of queries. 

It may be useful to leverage this popularity information in log mining and ranking studies, despite the absence of edge weights. For example, studies of query autocomplete \cite{Shokouhi:2013} have sometimes used the AOL and MSN logs mentioned in Section~\ref{sec:related}. Such studies need not only queries but also per-query popularity information. It would be possible to use the popularity information from Figure~\ref{fig:degree_distribution} to make the ORCAS queries usable in the same way.

ORCAS queries were selected based on connection to the TREC corpus, click aggregation, anonymity filtering and other filtering criteria. They were not otherwise selected to be natural language question answering queries. By contrast, the TREC DL queries were selected in the creation of the MS MARCO question answering task~\cite{bajaj2016ms}. This would lead us to expect that ORCAS queries and TREC DL queries have somewhat different characteristics.

Two such characteristics are query length and vocabulary. Figure~\ref{fig:orcas_terms} shows that the ORCAS queries tend to have fewer words than TREC DL queries. Table~\ref{tab:first word} compares the first word of queries. TREC DL queries are much more likely to have the word \emph{what} at the start of a query (39.4\% vs 3.5\%), although the ORCAS data has more \emph{what} queries in total (362K vs 145K). Both querysets have a healthy distribution of common and rare terms overall, we illustrate this for ORCAS data (also Figure~\ref{fig:orcas_terms}), but TREC DL has a similarly healthy vocabulary distribution. The differences in query length and the likelihood of having question words at the start mainly reflect the selection criteria of TREC DL queries, which are a subset of query traffic suitable for the MS MARCO question answering dataset, favoring natural language questions.

\section{Retrieval experiments}
\label{sec:experiment}
To study the effectiveness of the ORCAS data for training deep neural models, we conduct preliminary retrieval experiments on the document ranking task in the TREC 2019 Deep Learning Track~\citep{craswell2020overview}.
We describe these experiments and corresponding results here.

\paragraph{Data}
The TREC deep learning benchmark for document reranking provides a large training dataset containing 367,013 queries and 384,597 positively labeled query-document pairs from the MS MARCO dataset~\citep{bajaj2016ms}.
For every query, a set of $100$ documents is retrieved using Indri~\citep{strohman2005indri} and provided as part of the dataset.
Feature-based~\citep{Liu:2009} and representation learning-based~\citep{mitra2018introduction} learning to rank models are typically trained on this data by employing optimization objectives that contrast relevant and non-relevant documents for a given query.
Nonrelevant documents can be sampled either from the collection distribution or from a distribution that is more biased towards documents with at least partial matches with the query---\eg, the Indri top 100 retrieved results.
Previous work~\citep{mitra2017learning} has indicated that negative documents related to the query are more helpful for learning than negative documents sampled at random from the collection. 

Following a similar design to the MS MARCO training dataset, we generate a complementary dataset containing 10,405,342 queries and 18,823,602 positively labeled query-document pairs.
This is approximately $28$ times bigger than the MS MARCO training dataset in terms of the number of queries and approximately $49$ times bigger in terms of the number of positive labels.
To be consistent with the MS MARCO training data, we also provide the Indri top 100 retrieved results for each query.

\paragraph{Model}
We adopt a public implementation\footnote{\url{https://github.com/bmitra-msft/TREC-Deep-Learning-Quick-Start}} of a Transformer-based ranking model with query term independence~\citep{mitra2019incorporating} as our base architecture.
Both query and document terms are first encoded using a shared term embedding model.
The document term embeddings are then contextualized using stacked transformer layers.
We compute the cosine similarity between every pair of query and document term embeddings and then employ windowed kernle-pooling~\citep{hofstatter2019tu, hofstatter2020improving} and multiple feedforward layers to estimate the match between the document and each query term.
Finally, the scores are linearly combined across query terms.

\paragraph{Experiments}
We conduct two sets of experiments to study the usefulness of the ORCAS dataset
\begin{enumerate*}[label=(\roman*)]
    \item As a supplement to the MS MARCO training data, and
    \item As a document description field in addition to URL, title, and body.
\end{enumerate*}

For our first study, we compare the retrieval effectiveness of the Transformer-Kernel model~\citep{hofstatter2019tu} when trained on a combination of MS MARCO and ORCAS training data to training on MS MARCO data alone.
When the two datasets are combined, a two-step sampling is employed for training sample selection---we first randomly select one of the two training datasets with equal probability and then randomly sample a query from the selected dataset with uniform probability.
This means that our training model sees an equal proportion of samples from both datasets during training in spite of their significant difference in size.
This is done intentionally to control against diverging too far from the MS MARCO query distribution, since our test queries also come from MS MARCO.
We conduct two sets of experiments by sampling the negative document for training from
\begin{enumerate*}[label=(\roman*)]
    \item the collection distribution, and
    \item the Indri top-$100$ document distribution,
\end{enumerate*}
respectively.
For document representation we consider a maximum of the first $800$ terms.

In our second study, we use the ORCAS dataset to generate an additional document field and compare the performance of the base model trained on MS MARCO labels with and without the ORCAS field.
We restrict the document representation, across all fields, to $4000$ terms and the ORCAS field, specifically, to a maximum of $2000$ terms.
To handle the longer document text input we employ the Conformer-Kernel~\citep{mitra2020conformer} model as our base architecture for this study.

Across both studies, we evaluate our models on the 43 test queries from the 2019 edition of the track using the corresponding NIST labels provided as a reusable benchmark.
We report MRR and NDCG for each run.
The models are trained using the Adam optimizer and the RankNet objective~\citep{burges2005learning} with a learning rate of 0.0001.
All hyperparameters are consistent across different training runs.

\paragraph{Results}
Table~\ref{tab:retrieval-results1} summarizes the findings from the first study.
Under both negative sampling settings, we find that training on the combination of the two datasets gives roughly equivalent results to training on MS MARCO data only.
Although the ORCAS data has a higher number, the difference is not statistically significant on our 43 test queries.

\begin{table}[]
\centering
\caption{Retrieval experiment results on the document reranking task from the TREC Deep Learning benchmark to study the usefulness of the ORCAS dataset for traininig.}
\begin{tabular}{lll}
\toprule
Training data & MRR & NDCG \\
\midrule
\multicolumn{3}{l}{\textbf{Negatives sampled from full collection}} \\
MS MARCO only & $0.798$ & $0.505$ \\
MS MARCO + ORCAS & $\mathbf{0.807}$ & $\mathbf{0.509}$ \\
\midrule
\multicolumn{3}{l}{\textbf{Negatives sampled from top 100 candidates}} \\
MS MARCO only & $0.909$ & $0.574$ \\
MS MARCO + ORCAS & $\mathbf{0.924}$ & $\mathbf{0.582}$ \\
\bottomrule
\end{tabular}
\label{tab:retrieval-results1}
\end{table}

The results from our second study is presented in Table~\ref{tab:retrieval-results2}.
Similar to our first study we find that the inclusion of ORCAS dataset results in higher retrieval metrics but the difference is not statistically significant on our small test set.

\begin{table}[]
\centering
\caption{Retrieval experiment results on the document full-ranking task from the TREC Deep Learning benchmark to study the usefulness of the ORCAS dataset as an additional document field.}
\begin{tabular}{lll}
\toprule
Document fields & MRR & NDCG \\
\midrule
URL + Title + Body & $0.902$ & $0.616$ \\
URL + Title + ORCAS + Body & $\mathbf{0.931}$ & $\mathbf{0.629}$ \\
\bottomrule
\end{tabular}
\label{tab:retrieval-results2}
\end{table}

Although these initial studies did not yield significantly better results, we posit that models with larger number of layers or learnable parameters could perform better, taking advantage of the larger size of the ORCAS dataset. 

\section{Evaluating using ORCAS labels}
\label{sec:evaluation_results}

To analyze the viability of ORCAS query-URL pairs as positive relevance labels, we also used them in evaluation, to test the MRR of the $38$ runs in the 2019 TREC DL  document ranking task. For the 200 queries in the 2019 runs, we were able to identify 83 positive labels for 28 queries in the ORCAS dataset, whereas the official NIST labels were for 43 of the queries. 

\begin{figure}
    \centering
    \includegraphics[width=\linewidth]{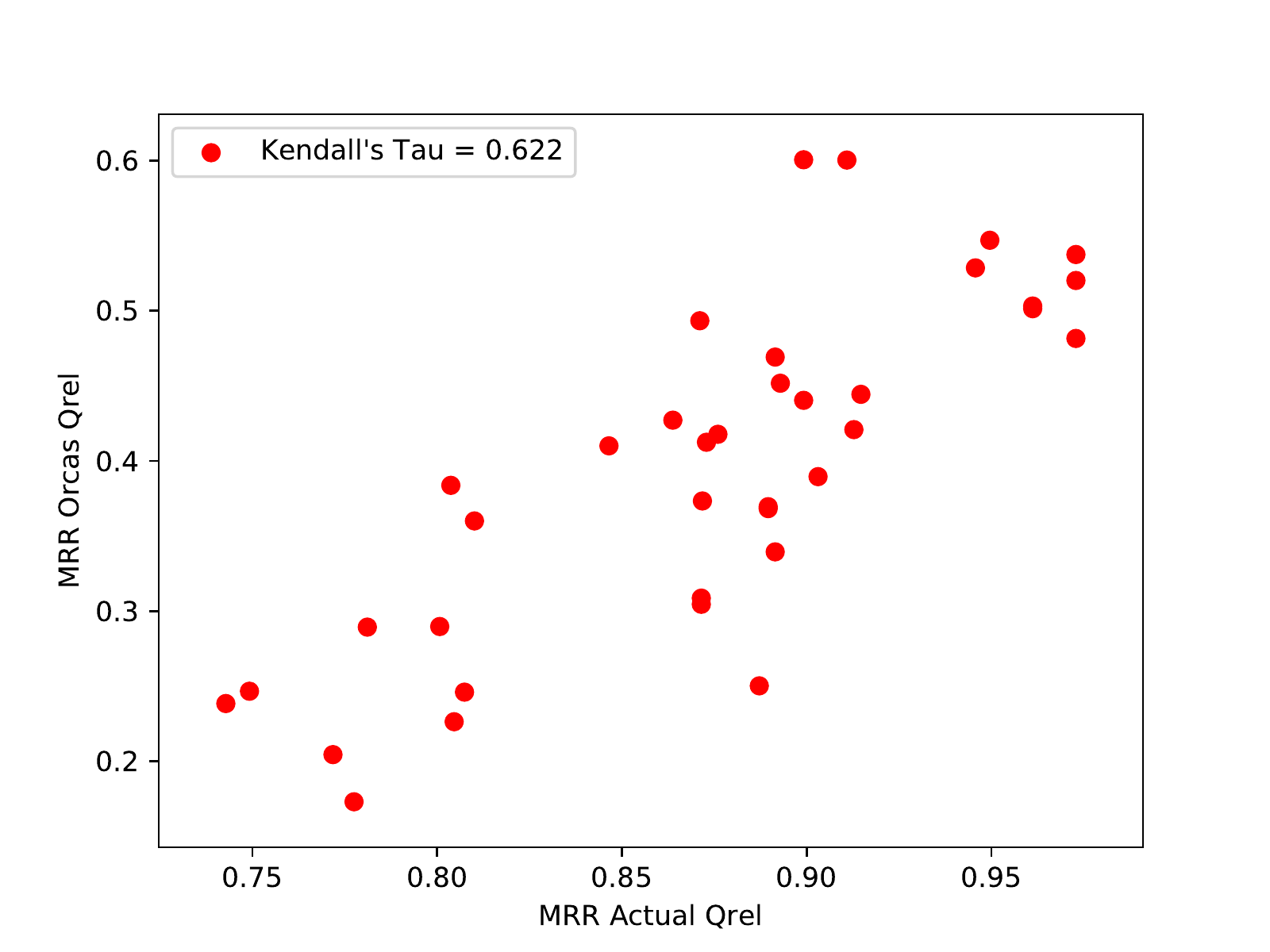}
    \caption{MRR values obtained when the runs submitted to the document ranking task of the TREC Deep Learning Track are evaluated with ORCAS labels vs.~actual Qrel from the track. }
    \label{eval_MRR}
\end{figure}

We then compared the evaluation results obtained using the subset from the ORCAS dataset with the evaluation results from the official test collection from TREC, using MRR as the evaluation metric. The comparison of the two evaluation results are shown in Figure~\ref{eval_MRR}. The figure reports also reports the Kendall's tau correlation between the rankings obtained using the two sets of metrics. Considering the small size of the ORCAS subset that is common with the TREC DL test collection, the correlations with the official MRR results look reasonable. These correlations would probably improve if all the queries in the ORCAS test data were to be used.

\section{Conclusion}
\label{sec:conclusion}

ORCAS is a click dataset with more than 10 million queries, connected to documents in the TREC Deep Learning Track document corpus. The queries can be released without privacy violation because they are not connected to any user ID or session ID, and only queries that were entered by many users are included. The dataset was also filtered, for example to exclude adult and offensive queries.

Compared to datasets used in previous log mining and ranking studies based on query-URL pairs, ORCAS is of comparable size to many of them, but is also an open dataset rather than a proprietary one. This could allow the replication and extension of a variety of past log mining studies and ranking studies.

We train a ranker using a mixture of ORCAS and TREC DL data, with our initial experiments showing some sign of promising results, but no statistically significant gains. Correctly weighting and adjusting the data to achieve a significant gain on a test set of 43 queries is left as future work, but may be possible given that the ORCAS training set is 49x larger than the TREC DL set. We also show that the ORCAS clicks can be used to calculate an MRR metric, evaluating TREC runs with a Tau correlation of 0.622 compared to MRR using TREC labels. This is further evidence that the ORCAS data can be used to augment labeled data, in training and evaluation.

The dataset could have several other uses. After mining to find related queries, the query pairs could futher be used for synonym mining, for use in query rewriting. Instead, the related queries could be issued separately, then a final result list constructed by blending the results. It would also be possible to mine the ORCAS query-document pairs to identify synonyms, viewing it as a translation problem, translating from query vocabulary to document vocabulary. Query-document connections can also be viewed as a kind of relevance feedback, for query expansion that relies on documents rather than synonym mining. The queries themselves could be used whenever a query histogram is useful, for simulating query autocompletion, or sampling queries for a future test collection. Since query-document connections are at the heart of information retrieval, we hope ORCAS proves a useful dataset for these and other future applications.




\balance
\bibliographystyle{ACM-Reference-Format}
\begin{scriptsize}
\bibliography{bibtex}
\end{scriptsize}

\end{document}